\documentclass[fleqn,usenatbib]{mnras}

\usepackage{newtxtext,newtxmath}

\usepackage[T1]{fontenc}
\usepackage{ae,aecompl}
\usepackage{graphicx}	
\usepackage{amsmath}	
\usepackage{amssymb}	
\usepackage{natbib,longtable,times,multicol}
\usepackage{caption}
\usepackage[caption = false]{subfig}
\newcommand{\todo}{\ifmmode \text{\Huge{\(\bullet\)}} \else {\Huge$\bullet$}\fi}
\newcommand{\tido}{\ifmmode {\bullet} \else $\bullet$\fi}

\newcommand{\E        }[1]{\ifmmode 10^{#1} \else $10^{#1}$\fi}
\newcommand{\tE        }[1]{\ifmmode \times10^{#1} \else $\times10^{#1}$\fi}
\newcommand{\til}{\ifmmode \sim \else $\sim$\fi}
\renewcommand{\~} {\ifmmode \sim \else $\sim$\fi}

\newcommand{\pc}	{\ifmmode {\rm pc} \else pc\fi}
\newcommand{\ld}	{\ifmmode {\rm l.d.} \else l.d.\fi}
\newcommand{\kms}	{\ifmmode {\rm km\,s}^{-1} \else km\,s$^{-1}$\fi}
\newcommand{\cc}	{\ifmmode {\rm cm}^{-3}    \else cm$^{-3}$\fi}
\newcommand{\cmii}	{\ifmmode {\rm cm}^{-2}    \else cm$^{-2}$\fi}
\newcommand{\ergs}	{\ifmmode {\rm erg\,s}^{-1} \else erg s$^{-1}$\fi}
\newcommand{\ergcms}	{\ifmmode {\rm erg\,cm}^{-2}\,{\rm s}^{-1} \else erg\,cm$^{-2}$\,s$^{-1}$\fi}
\newcommand{\ergcmsA}	{\ifmmode {\rm erg\,cm}^{-2}\,{\rm s}^{-1}\,{\rm\AA}^{-1}
\else erg\,cm$^{-2}$\,s$^{-1}$\,\AA$^{-1}$\fi}
\newcommand{  \ergcmsHz  }{\ifmmode{\rm erg\,cm}^{-2}\,{\rm s}^{-1}\,{\rm Hz}^{-1}
                       \else ergs\,cm$^{-2}$\,s$^{-1}$\,Hz$^{-1}$\fi}
\newcommand{\kev}	{\ifmmode {\rm keV} \else keV\fi}

\newcommand{\mic}	{\ifmmode {\rm \mu m} \else $\mu$m\fi}
\newcommand{\vFWHM}	{\ifmmode v_{\mbox{\tiny FWHM}} \else $v_{\mbox{\tiny FWHM}}$\fi}
\newcommand{\vBLR}	{\ifmmode v_{\mbox{\tiny BLR}} \else $v_{\mbox{\tiny BLR}}$\fi}
\newcommand{\sigBLR}	{\ifmmode \sigma_{\mbox{\tiny BLR}} \else $\sigma_{\mbox{\tiny BLR}}$\fi}
\newcommand{\vNLR}	{\ifmmode v_{\mbox{\tiny NLR}} \else $v_{\mbox{\tiny NLR}}$\fi}
\newcommand{\tauBLR}	{\ifmmode \tau_{\mbox{\tiny BLR}} \else $\tau_{\mbox{\tiny BLR}}$\fi}

\newcommand{\Hubble}	{\ifmmode {\rm km\,s}^{-1}\,{\rm Mpc}^{-1} \else km\,s$^{-1}$\,Mpc$^{-1}$\fi}
\newcommand{\NDunit}	{\ifmmode {\rm Mpc}^{-3} \else Mpc$^{-3}$\fi}
\newcommand{\LFunit}	{\ifmmode {\rm Mpc}^{-3}\,{\rm mag}^{-1} \else Mpc$^{-3}$\,mag$^{-1}$\fi}
\newcommand{\MFunit}	{\ifmmode {\rm Mpc}^{-3}\,{\rm dex}^{-1} \else Mpc$^{-3}$\,dex$^{-1}$\fi}

\newcommand{\Msun}{\ifmmode M_{\odot} \else $M_{\odot}$\fi}
\newcommand{\Lsun}{\ifmmode L_{\odot} \else $L_{\odot}$\fi}
\newcommand{\Zsun}{\ifmmode Z_{\odot} \else $Z_{\odot}$\fi}
\newcommand{\mpyr}{\ifmmode \Msun\,{\rm yr}^{-1} \else $\Msun\,{\rm yr}^{-1}$\fi}

\newcommand{\qnote}{\ifmmode q_{0} \else $q_{0}$\fi}
\newcommand{\Hnote}{\ifmmode H_{0} \else $H_{0}$\fi}
\newcommand{\hnote}{\ifmmode h_{0} \else $h_{0}$\fi}
\newcommand{\anote}{\ifmmode a_{0} \else $a_{0}$\fi}
\newcommand{\tnote}{\ifmmode t_{0} \else $t_{0}$\fi}

\newcommand{  \Halpha   }{\ifmmode {\rm H}\alpha \else H$\alpha$\fi}
\newcommand{  \ha   	}{\ifmmode {\rm H}\alpha \else H$\alpha$\fi}
\newcommand{  \Hbeta    }{\ifmmode {\rm H}\beta \else H$\beta$\fi}
\newcommand{  \hb    	}{\ifmmode {\rm H}\beta \else H$\beta$\fi}
\newcommand{  \Hgamma   }{\ifmmode {\rm H}\gamma \else H$\gamma$\fi}
\newcommand{  \Hdelta   }{\ifmmode {\rm H}\delta \else H$\delta$\fi}
\newcommand{  \Lya      }{\ifmmode {\rm Ly}\alpha \else Ly$\alpha$\fi}
\newcommand{  \Lyb      }{\ifmmode {\rm Ly}\beta \else Ly$\beta$\fi}
\newcommand{  \Pa       }{\ifmmode {\rm P}\alpha \else P$\alpha$\fi}
\newcommand{  \Pb       }{\ifmmode {\rm P}\beta \else P$\beta$\fi}
\newcommand{  \Bra      }{\ifmmode {\rm Br}\alpha \else Br$\alpha$\fi}
\newcommand{  \Brg      }{\ifmmode {\rm Br}\gamma \else Br$\gamma$\fi}
\newcommand{  \hii      }{\ifmmode {\rm H}\,\textsc{ii} \else H\,\textsc{ii}\fi}
\newcommand{  \hei      }{\ifmmode {\rm He}\,\textsc{i} \else He\,\textsc{i}\fi}
\newcommand{  \heii     }{\ifmmode {\rm He}\,\textsc{ii} \else He\,\textsc{ii}\fi}
\newcommand{  \HeIIuv   }{\ifmmode {\rm He}\,\textsc{ii}\,\lambda1640 \else He\,\textsc{ii}\,$\lambda1640$\fi}
\newcommand{  \HeIIop   }{\ifmmode {\rm He}\,\textsc{ii}\,\lambda4686 \else He\,\textsc{ii}\,$\lambda4686$\fi}
\newcommand{  \cii      }{\ifmmode {\rm C}\,\textsc{ii}  \else C\,\textsc{ii}\fi}
\newcommand{  \ciii     }{\ifmmode {\rm C}\,\textsc{iii}\right] \else C\,\textsc{iii}]\fi}
\newcommand{  \CIII     }{\ifmmode {\rm C}\,\textsc{iii}\right]\,\lambda1909 \else C\,\textsc{iii}]\,$\lambda1909$\fi}
\newcommand{  \civ      }{\ifmmode {\rm C}\,\textsc{iv}  \else C\,\textsc{iv}\fi}
\newcommand{  \CIV      }{\ifmmode {\rm C}\,\textsc{iv}\,\lambda1549 \else C\,\textsc{iv}\,$\lambda1549$\fi}
\newcommand{  \nii      }{\ifmmode [{\rm N}\,\textsc{ii}]  \else [N\,\textsc{ii}]\fi}
\newcommand{  \niii     }{\ifmmode {\rm N}\,\textsc{iii} \else N\,\textsc{iii}\fi}
\newcommand{  \niv      }{\ifmmode {\rm N}\,\textsc{iv}  \else N\,\textsc{iv}\fi}
\newcommand{  \NIVuv    }{\ifmmode {\rm N}\,\textsc{iv}\,\lambda1486 \else N\,\textsc{iv}\,$\lambda1486$\fi}
\newcommand{  \nv       }{\ifmmode {\rm N}\,\textsc{v}   \else N\,\textsc{v}\fi}
\newcommand{\oi}{\ifmmode \left[{\rm O}\,\textsc{i}\right] \else [O\,{\sc i}]\fi}
\newcommand{\OI}{\ifmmode \left[{\rm O}\,\textsc{i}\right]\,\lambda6300 \else [O\,{\sc i}]$\,\lambda6300$\fi}

\newcommand{\oii}{\ifmmode \left[{\rm O}\,\textsc{ii}\right] \else [O\,{\sc ii}]\fi}
\newcommand{\OII}{\ifmmode \left[{\rm O}\,\textsc{ii}\right]\,\lambda3727 \else [O\,{\sc ii}]\,$\lambda3727$\fi}
\newcommand{\oiii}{\ifmmode \left[{\rm O}\,\textsc{iii}\right] \else [O\,{\sc iii}]\fi}
\newcommand{\OIII}{\ifmmode \left[{\rm O}\,\textsc{iii}\right]\,\lambda5007 \else [O\,{\sc iii}]\,$\lambda5007$\fi}

\newcommand{\NII}{\ifmmode \left[{\rm N}\,\textsc{ii}\right]\,\lambda6583 \else [N\,{\sc ii}]$\,\lambda6583$\fi}

\newcommand{\NeIII}{\ifmmode \left[{\rm Ne}\,\textsc{iii}\right]\,\lambda3968 \else [Ne\,{\sc iii}]$\,\lambda3968$\fi}

\newcommand{\NeV}{\ifmmode \left[{\rm Ne}\,\textsc{v}\right]\,\lambda3426 \else [Ne\,{\sc v}]$\,\lambda3426$\fi}

\newcommand{\HeII}{\ifmmode {\rm He}\,\textsc{ii}\,\lambda4686 \else He\,{\sc ii}$\,\lambda4686$\fi}

\newcommand{\sii}{\ifmmode \left[{\rm S}\,\textsc{ii}\right] \else [S\,{\sc ii}]\fi}

\newcommand{\SII}{\ifmmode \left[{\rm S}\,\textsc{ii}\right]\,\lambda6717,6731 \else [S\,{\sc ii}]$\,\lambda6717,6731$\fi}

\newcommand{  \OIIIuv   }{\ifmmode {\rm O}\,\textsc{iii}\,\lambda1663 \else O\,\textsc{iii}\,$\lambda1663$\fi}
\newcommand{  \oiv      }{\ifmmode {\rm O}\,\textsc{iv}  \else O\,\textsc{iv}\fi}
\newcommand{  \OIVuv    }{\ifmmode {\rm O}\,\textsc{iv}\,\lambda1402  \else O\,\textsc{iv}\,$\lambda1402$\fi}
\newcommand{  \OIVIR    }{\ifmmode {\rm O}\,\textsc{iv}\,25.9\,\mu {\rm m} \else O\,\textsc{iv}\,$25.9\,\mu$m\fi}
\newcommand{  \ovi      }{\ifmmode {\rm O}\,\textsc{vi}   \else O\,\textsc{vi}\fi}
\newcommand{  \Ovi      }{\ifmmode {\rm O}\,\textsc{vi}\,\lambda1035 \else O\,\textsc{vi}\,$\lambda1035$\fi}
\newcommand{  \nei      }{\ifmmode {\rm Ne}\,\textsc{i}   \else Ne\,\textsc{i}\fi}
\newcommand{  \neii     }{\ifmmode {\rm Ne}\,\textsc{ii}  \else Ne\,\textsc{ii}\fi}
\newcommand{  \NeiiIR   }{\ifmmode {\rm Ne}\,\textsc{ii}\,12.8\,\mu {\rm m} \else Ne\,\textsc{ii}\,$12.8\,\mu$m\fi}
\newcommand{  \neiii    }{\ifmmode {\rm Ne}\,\textsc{iii} \else Ne\,\textsc{iii}\fi}
\newcommand{  \neiv     }{\ifmmode {\rm Ne}\,\textsc{iv}  \else Ne\,\textsc{iv}\fi}
\newcommand{  \nev      }{\ifmmode {\rm Ne}\,\textsc{v}   \else Ne\,\textsc{v}\fi}
\newcommand{  \NevIR    }{\ifmmode {\rm Ne}\,\textsc{v}\,24.3\,\mu {\rm m} \else Ne\,\textsc{v}\,$24.3\,\mu$m\fi}
\newcommand{  \nevi     }{\ifmmode {\rm Ne}\,\textsc{vi}  \else Ne\,\textsc{vi}\fi}
\newcommand{  \mgi      }{\ifmmode {\rm Mg}\,\textsc{i}   \else Mg\,\textsc{i}\fi}
\newcommand{  \mgii     }{\ifmmode {\rm Mg}\,\textsc{ii}  \else Mg\,\textsc{ii}\fi}
\newcommand{  \MgII     }{\ifmmode {\rm Mg}\,\textsc{ii}\,\lambda2798 \else Mg\,\textsc{ii}\,$\lambda2798$\fi}

\newcommand{  \siii     }{\ifmmode {\rm S}\,\textsc{iii} \else S\,\textsc{iii}\fi}
\newcommand{  \siv      }{\ifmmode {\rm S}\,\textsc{iv}  \else S\,\textsc{iv}\fi}
\newcommand{  \sili     }{\ifmmode {\rm Si}\,\textsc{i}   \else Si\,\textsc{i}\fi}
\newcommand{  \silii    }{\ifmmode {\rm Si}\,\textsc{ii}  \else Si\,\textsc{ii}\fi}
\newcommand{  \Siliv    }{\ifmmode {\rm Si}\,\textsc{iv}  \else Si\,\textsc{iv}\fi}
\newcommand{  \SilIVuv  }{\ifmmode {\rm Si}\,\textsc{iv}\,\lambda1400  \else Si\,\textsc{iv}\,$\lambda1400$\fi}

\newcommand{  \caii     }{\ifmmode {\rm Ca}\,\textsc{ii}   \else Ca\,\textsc{ii}\fi}
\newcommand{  \feii     }{\ifmmode {\rm Fe}\,\textsc{ii}  \else Fe\,\textsc{ii}\fi}
\newcommand{  \feiii    }{\ifmmode {\rm Fe}\,\textsc{iii} \else Fe\,\textsc{iii}\fi}

\newcommand{ \Lhb   }{\ifmmode L\left(\hb\right) \else $L\left(\hb\right)$\fi}
\newcommand{ \fwhb  }{\ifmmode {\rm FWHM}\left(\hb\right) \else FWHM(\hb)\fi}
\newcommand{ \Lha   }{\ifmmode L\left(\ha\right) \else $L\left(\ha\right)$\fi}
\newcommand{ \fwha  }{\ifmmode {\rm FWHM}\left(\ha\right) \else FWHM(\ha)\fi}
\newcommand{ \Lmg   }{\ifmmode L\left(\mgii\right) \else $L\left(\mgii\right)$\fi}
\newcommand{ \fwmg  }{\ifmmode {\rm FWHM}\left(\mgii\right) \else FWHM(\mgii)\fi}
\newcommand{ \Lciv  }{\ifmmode L\left(\civ\right) \else $L\left(\civ\right)$\fi}
\newcommand{ \fwciv }{\ifmmode {\rm FWHM}\left(\civ\right) \else FWHM(\civ)\fi}
\newcommand{ \fwhm  }{\ifmmode {\rm FWHM} \else FWHM\fi} 
\newcommand{ \voff  }{\ifmmode v_{\rm off} \else $v_{\rm off}$\fi} 

\newcommand{ \mumg  }{\ifmmode \mu\left(\mgii\right) \else $\mu\left(\mgii\right)$\fi}
\newcommand{ \fmg   }{\ifmmode f\left(\mgii\right) \else $f\left(\mgii\right)$\fi}
\newcommand{ \muciv }{\ifmmode \mu\left(\civ\right) \else $\mu\left(\civ\right)$\fi}
\newcommand{ \fciv  }{\ifmmode f\left(\civ\right) \else $f\left(\civ\right)$\fi}

\newcommand{  \auvo     }{\ifmmode \alpha_{\nu,{\rm UVO}} \else $\alpha_{\nu,{\rm UVO}}$\fi}
\newcommand{  \Ledd     }{\ifmmode L_{\rm Edd} \else $L_{\rm Edd}$\fi}
\newcommand{  \lamLlam  }{\ifmmode \lambda L_{\lambda} \else $\lambda L_{\lambda}$\fi}
\newcommand{  \lLl      }{\ifmmode \lambda L_{\lambda} \else $\lambda L_{\lambda}$\fi}
\newcommand{  \nuLnu    }{\ifmmode \nu L_{\nu} \else $\nu L_{\nu}$\fi}
\newcommand{  \nLn      }{\ifmmode \nu L_{\nu} \else $\nu L_{\nu}$\fi}
\newcommand{  \Luv      }{\ifmmode L_{1450} \else $L_{1450}$\fi}
\newcommand{  \Lop      }{\ifmmode L_{5100} \else $L_{5100}$\fi}
\newcommand{  \lLop     }{\ifmmode \log\left(\Lop/\ergs\right) \else $\log\left(\Lop/\ergs\right)$\fi}
\newcommand{  \Lthree   }{\ifmmode L_{3000} \else $L_{3000}$\fi}
\newcommand{  \lLthree  }{\ifmmode \log\left(\Lthree/\ergs\right) \else $\log\left(\Lthree/\ergs\right)$\fi}

\newcommand{\Fthree}{\ifmmode F_{3000} \else $F_{3000}$\fi}
\newcommand{\fuv}{\ifmmode f_{\lambda}\left(1450{\rm \AA}\right) \else $f_{\lambda}\left(1450 {\rm \AA}\right)$\fi}
\newcommand{\fthree}{\ifmmode f_{\lambda}\left(3000{\rm \AA}\right) \else $f_{\lambda}\left(3000{\rm \AA}\right)$\fi}
\newcommand{\fH}{\ifmmode f_{\lambda}\left(1.65\micron\right) \else
$f_{\lambda}\left(1.65\micron\right)$\fi}

\newcommand{\fbol}{\ifmmode f_{\rm bol} \else $f_{\rm bol}$\fi}
\newcommand{\fbolwv}{\ifmmode f_{\rm bol}\left(\lambda\right) \else $f_{\rm bol}\left(\lambda\right)$\fi}
\newcommand{\fbolopt}{\ifmmode f_{\rm bol}\left(5100{\rm \AA}\right) \else $f_{\rm bol}\left(5100{\rm \AA}\right)$\fi}
\newcommand{\fbolthree}{\ifmmode f_{\rm bol}\left(3000{\rm \AA}\right) \else $f_{\rm bol}\left(3000{\rm \AA}\right)$\fi}
\newcommand{\fboluv}{\ifmmode f_{\rm bol}\left(1450{\rm \AA}\right) \else $f_{\rm bol}\left(1450{\rm \AA}\right)$\fi}

\newcommand{  \mbh      }{\ifmmode M_{\rm BH} \else $M_{\rm BH}$\fi}
\newcommand{  \lmbh     }{\ifmmode \log\left(\mbh/\Msun\right) \else $\log\left(\mbh/\Msun\right)$\fi} 
\newcommand{  \lledd    }{\ifmmode L/L_{\rm Edd} \else $L/L_{\rm Edd}$\fi}
\newcommand{  \Lbol     }{\ifmmode L_{\rm bol} \else $L_{\rm bol}$\fi}
\newcommand{  \lbol     }{\ifmmode L_{\rm bol} \else $L_{\rm bol}$\fi}
\newcommand{  \lLbol    }{\ifmmode \log\left(\Lbol/\ergs\right) \else $\log\left(\Lbol/\ergs\right)$\fi} 
\newcommand{  \Lagn     }{\ifmmode L_{\rm AGN} \else $L_{\rm AGN}$\fi}
\newcommand{  \lagn     }{\ifmmode L_{\rm AGN} \else $L_{\rm AGN}$\fi}

\newcommand{  \tgrow     }{\ifmmode t_{\rm growth} \else $t_{\rm growth}$\fi}
\newcommand{  \tUni      }{\ifmmode t_{\rm Universe} \else $t_{\rm Universe}$\fi}

\newcommand{  \Mindot	}{\ifmmode \dot{M}_{\rm infall} \else $\dot{M}_{\rm infall}$\fi}
\newcommand{  \Mbhdot	}{\ifmmode \dot{M}_{\rm BH} \else $\dot{M}_{\rm BH}$\fi}
\newcommand{  \Maddot	}{\ifmmode \dot{M}_{\rm AD} \else $\dot{M}_{\rm AD}$\fi}

\newcommand{  \as	}{\ifmmode a_{\rm *} 		\else $a_{\rm *}$\fi}
\newcommand{  \avec	}{\ifmmode \vec{a}_{\rm *} 	\else $\vec{a}_{\rm *}$\fi}
\newcommand{  \re	}{\ifmmode \eta      	\else $\eta$\fi}

\newcommand{  \mseed    }{\ifmmode M_{\rm seed} \else $M_{\rm seed}$\fi}
\newcommand{  \mbul     }{\ifmmode M_{\rm Bulge} \else $M_{\rm Bulge}$\fi} 
\newcommand{  \mstar    }{\ifmmode M_{*} \else $M_{*}$\fi} 
\newcommand{  \mgal     }{\ifmmode M_{*} \else $M_{*}$\fi} 
\newcommand{  \mhost    }{\ifmmode M_{\rm Host} \else $M_{\rm Host}$\fi}
\newcommand{  \mm       }{\ifmmode M_{*}/M_{\rm BH} \else $M_{*}/M_{\rm BH}$\fi}
\newcommand{  \mmsmall  }{\ifmmode M_{\rm BH}/M_{*} \else $M_{\rm BH}/M_{*}$\fi}
\newcommand{  \mmlarge  }{\ifmmode M_{*}/M_{\rm BH} \else $M_{*}/M_{\rm BH}$\fi}
\newcommand{  \mmwp     }{\ifmmode \left(M_{*}/M_{\rm BH}\right) \else $\left(M_{*}/M_{\rm BH}\right)$\fi}
\newcommand{  \ml       }{\ifmmode M_{*}/L_{*} \else $M_{*}/L_{*}$\fi}
\newcommand{  \mlwp     }{\ifmmode \left(M_{*}/L\right) \else $\left(M_{*}/L\right)$\fi}
\newcommand{  \mlk      }{\ifmmode \left(M_{*}/L_{K}\right) \else $\left(M_{*}/L_{K}\right)$\fi}
\newcommand{  \sigs     }{\ifmmode \sigma_{*} \else $\sigma_{*}$\fi}
\newcommand{  \Reff     }{\ifmmode R_{\rm e} \else $R_{\rm e}$\fi}

\def \swiftbat {{\em Swift}/BAT\ }

\def \chandra {{\em Chandra\ }}
\def \chandrash {{\em Chandra}}

\newcommand{\bj}{\ifmmode b_{\rm J} \else $b_{\rm J}$\fi}

\newcommand{\iab}{\ifmmode i_{\rm AB} \else $i_{\rm AB}$\fi}

\newcommand{\jab}{\ifmmode J_{\rm AB} \else $J_{\rm AB}$\fi}
\newcommand{\hab}{\ifmmode H_{\rm AB} \else $H_{\rm AB}$\fi}
\newcommand{\kab}{\ifmmode K_{\rm AB} \else $K_{\rm AB}$\fi}

\newcommand{\jveg}{\ifmmode J_{\rm Vega} \else $J_{\rm Vega}$\fi}
\newcommand{\hveg}{\ifmmode H_{\rm Vega} \else $H_{\rm Vega}$\fi}
\newcommand{\kveg}{\ifmmode K_{\rm Vega} \else $K_{\rm Vega}$\fi}

\newcommand{  \Chisq    }{\ifmmode \chi^{2} \else $\chi^{2}$}
\newcommand{  \nelec    }{\ifmmode n_{e} \else $n_{e}$\fi}     
\newcommand{  \nh       }{\ifmmode n_{H} \else $n_{H}$\fi}     
\newcommand{  \Ncol     }{\ifmmode N_{col} \else $N_{col}$\fi} 
\newcommand{  \NH       }{\ifmmode N_{H} \else $N_{\rm H}$\fi}     

\def\ion#1#2{#1$\;${\small\rm\@Roman{#2}}\relax}

\newcommand{\OIIIa}{\ifmmode \left[{\rm O}\,\textsc{iii}\right]\,\lambda4959 \else [O\,{\sc iii}]\,$\lambda4959$\fi}
\newcommand{\NIIa}{\ifmmode \left[{\rm N}\,\textsc{ii}\right]\,\lambda6548 \else [N\,{\sc ii}]\,$\lambda6548$\fi}
\newcommand{\SIIa}{\ifmmode \left[{\rm S}\,\textsc{ii}\right]\,\lambda6716 \else [S\,{\sc ii}]\,$\lambda6716$\fi}
\newcommand{\SIIb}{\ifmmode \left[{\rm S}\,\textsc{ii}\right]\,\lambda6732 \else [S\,{\sc ii}]\,$\lambda6731$\fi}
\newcommand{\NeVa}{\ifmmode \left[{\rm Ne}\,\textsc{v}\right]\,\lambda3346 \else [Ne\,{\sc v}]\,$\lambda3346$\fi}
\newcommand{\NeVb}{\ifmmode \left[{\rm Ne}\,\textsc{v}\right]\,\lambda3426 \else [Ne\,{\sc v}]\,$\lambda3426$\fi}
\newcommand{\NeIIIa}{\ifmmode \left[{\rm Ne}\,\textsc{iii}\right]\,\lambda3869 \else [Ne\,{\sc iii}]\,$\lambda3869$\fi}
\newcommand{\NeIIIb}{\ifmmode \left[{\rm Ne}\,\textsc{iii}\right]\,\lambda3968 \else [Ne\,{\sc iii}]\,$\lambda3968$\fi}
\newcommand{\Mgb}{\ifmmode \left{\rm Mg}\,\textsc{i}\right\,\lambda5175 \else Mg\,{\sc i}\,$\lambda5175$\fi}

\newcommand{\mgb}{\ifmmode \left{\rm Mg}\,\textsc{i}\right \else Mg\,{\sc i}\fi}

\newcommand{\Cahk}{\ifmmode \left[{\rm Ca H+K}\,\textsc{ii}\right\,\lambda3935,3968 \else Ca H+K$\,\lambda3935,3968$\fi}

\def\Lsun{\hbox{$\rm\thinspace L_{\odot}$}}

\def\pc{{\rm\thinspace pc}}

\def \mytorus {{\tt MYtorus\ }}

\newcommand {\nhunit} {cm$^{-2}$}
\newcommand {\feka} {Fe~K$\alpha$}

\title[Inferring CT AGN candidates using the spectral curvature]{Inferring Compton-thick AGN candidates at z>2 with Chandra using the >8 keV restframe spectral curvature}

\author[]{
 \parbox[t]{18cm}{L. Baronchelli$^1$, M. Koss$^1$, K. Schawinski$^1$, C. Cardamone$^2$, F. Civano$^{3,4}$, A. Comastri$^{5}$, M. Elvis$^{4}$, G. Lanzuisi$^{5,6}$, S. Marchesi$^{5,7}$, C. Ricci$^{8,9}$, M. Salvato$^{10}$, B. Trakhtenbrot$^{1}$, E. Treister$^{8}$}\\\\
$^{1}$Institute for Astronomy, Department of Physics, ETH Zurich, Wolfgang-Pauli-Strasse 27, CH-8093 Zurich, Switzerland\\
$^{2}$Math $\&$ Science Department, Wheelock College, 200 The Riverway, Boston, MA 02215, USA\\
$^{3}$Yale Center for Astronomy and Astrophysics, 260 Whitney Avenue, New Haven, CT 06520, USA\\
$^{4}$Harvard Smithsonian Center for Astrophysics, 60 Garden Street, Cambridge, MA 02138, USA\\
$^{5}$ INAF - Osservatorio Astronomico di Bologna, via Ranzani 1, 40127 Bologna, Italy\\
$^{6}$Dipartimento di Fisica e Astronomia, Alma Mater Studiorum, Universit\`{a} di Bologna, viale Berti Pichat 6/2, 40127, Bologna, Italy\\
$^{7}$Department  of  Physics  $\&$  Astronomy,  Clemson  University, Clemson, SC 29634, USA\\
$^{8}$Instituto de Astrofisica, Facultad de Fisica, Pontificia Universidad Catolica de Chile, Casilla 306, Santiago 22, Chile\\
$^{9}$Kavli Institute for Astronomy and Astrophysics, Peking University, Beijing 100871, China\\
$^{10}$Max-Planck-Institut f\"{u}r extraterrestrische Physik, Giessenbachstrasse 1, D85748 Garching bei M\"{u}nchen, Germany\\
}

\date{Accepted 2017 June 20 . Received 2017 June 20; in original form 2016 November 23}

\pubyear{2016}

\begin{document}
\label{firstpage}
\pagerange{\pageref{firstpage}--\pageref{lastpage}}
\maketitle

\begin{abstract}
To fully understand cosmic black hole growth we need to constrain the population of heavily obscured active galactic nuclei (AGN) at the peak of cosmic black hole growth ($z\sim$1--3). Sources with obscuring column densities higher than $\mathrm{10^{24}\,atoms\,cm^{-2}}$, called Compton-thick (CT) AGN, can be identified by excess X-ray emission at $\sim 20$--$30$ keV, called the "Compton hump". We apply the recently developed Spectral Curvature (SC) method to high-redshift AGN ($2<z<5$) detected with \textit{Chandra}. This method parametrizes the characteristic "Compton hump" feature cosmologically redshifted into the X-ray band at observed energies $<10$ keV. We find good agreement in CT AGN found using the SC method and bright sources fit using their full spectrum with X-ray spectroscopy. In the \textit{Chandra} deep field south, we measure a CT fraction of 17$^{+19}_{-11}\%$ (3/17) for sources with observed luminosity $>5\times 10^{43} \mathrm{erg\,s^{-1}}$. In the Cosmological evolution survey (COSMOS), we find an observed CT fraction of $15^{+4}_{-3}\%$ (40/272) or $32\pm11 \%$ when corrected for the survey sensitivity. When comparing to low redshift AGN with similar X-ray luminosities, our results imply the CT AGN fraction is consistent with having no redshift evolution.  Finally, we provide SC equations that can be used to find high-redshift CT AGN ($z>1$) for current (\textit{XMM-Newton}) and future  (\textit{eROSITA} and \textit{ATHENA}) X-ray missions.

\end{abstract}




\section{Introduction}
\label{sec:intro}

Active galactic nuclei (AGN) are believed to be powered during accretion episodes in which matter from galactic scales is accreted onto the central supermassive black hole (SMBH) \citep[e.g.,][]{Soltan1982, Marconi2004,Merloni2006}. During these accretion phases, periods of maximal growth occur in the SMBH \citep[e.g.,][]{FerrareseSpaceSci.Rev.116:523-6242005,Johnson2013}. Due to the large amount of matter involved during the accretion of a SMBH, a significant fraction of AGN is obscured from sight \citep[e.g.,][]{Balokovic2014,Brightman2014}.
Thus, to understand the evolution history of all the SMBHs through cosmic time, we need a complete census of the AGN population including the heavily obscured sources \citep[e.g.,][]{Treister2009,Ueda2014,Ricci2015,Buchner2015}. 
The capability of the X-ray emission at energies >10 keV to penetrate obscuring matter makes them one of the best tools to study obscured AGN \citep{Risaliti1999,Barger2003,Georgantopoulos2009}. The detection of AGN can, however, become very challenging when the absorption reaches Compton-thick (CT) levels \citep{ Georgantopoulos2010, Lanzuisi2015, Brandt2015}. We define an AGN as CT when it is surrounded by obscuring material with column density on the line of sight larger than the inverse Thomson cross-section \citep[$\mathrm{N_H \geq \sigma_T^{-1} \approx 1.5 \times 10^{24}\, atoms\, cm^{-2},}$][]{Comastri2016}.

The study of highly obscured sources, such as CT AGN, is crucial to achieve a complete census of the accreting SMBH population and to obtain an unbiased X-ray luminosity function  \citep[e.g.,][]{Fabian1999,Alexander2007,Georgakakis2015}.    \cite{Gilli2007}  found that to explain the cosmic X-ray background (XRB) peak at $\sim 30$ keV the fraction of CT AGN must be equivalent to the fraction of moderately obscured sources (21 < $\mathrm{log(N_H)}$ < 24). Their results agree with \cite{Fiore2008}. 
\cite{Daddi2007} studied the population at $z \sim 2$ showing an excess in the mid-IR wavelength suggesting a space density of CT AGN of $\sim 2.6 \times 10^{-4}$ Mpc$^{-3}$. However, they found that even if the population of CT AGN has a large space density, the CT contribution to the still missing XRB is of the order of $10\% $--$ 25\%$. This result is consistent with what has been found by \cite{Treister2009a}.  The analysis of the hard X-ray luminosity function from \cite{Ueda2014} using X-ray data from \textit{Swift/BAT}, \textit{MAXI}, \textit{ASCA}, \textit{XMM-Newton}, \textit{Chandra} and \textit{ROSAT} reveals that the number of sources with column density between $\mathrm{log(N_H)}=24$--$25$ must be equal to the number of sources with $\mathrm{log(N_H)}=23$--$24$ to explain the cosmic XRB emission at 20 keV. This result is similar to what has been found by \cite{Gilli2007}. X-ray spectral analysis of the 4 Ms \textit{Chandra} Deep Field South (CDF-S) by \cite{Brightman2012} using spectral models from \cite{Brightman2011} showed a CT fraction in the nearby Universe of $\sim 20\%$ growing to $\sim 40\%$ at redshift $z=1$--$4$.
However, \cite{Buchner2015} combined deep and wide-area \textit{Chandra} and \textit{XMM-Newton} X-ray surveys and they did not find any evidence of the redshift evolution of the CT fraction which they found to be $38^{+8}_{-7}\%$ on a redshift range from 0.5 to 2. This could be explained by the difference in the analyzed luminosity ranges as the sample in \cite{Buchner2015} includes sources with X-ray luminosities down to $10^{43.2}$ erg $\mathrm{s^{-1}}$. \cite{Ricci2015} found the CT fraction to be luminosity dependent with $32\pm 7\%$ at luminosities $\mathrm{log(L_{14-195 keV})=}$40--43.7, while only $21 \pm 5\%$ at higher luminosities  $\mathrm{log(L_{14-195 keV})=}$43.7--46. This result is similar to what found by \cite{Civano2015} who performed an analysis of the Cosmological Evolution Survey (COSMOS) field with \textit{NuSTAR}, finding a CT fraction between 13$\%$ and 20$\%$ at redshift z=0.04--2.5. However, the result of \cite{Ricci2015} is corrected from bias, while the fraction in \cite{Civano2015} is not. We note that some studies have suggested that most of them "missing" XRB is expected to be produced by objects with intrinsic luminosity smaller than $\mathrm{10^{44}\,erg\,s^{-1}}$ and z < 1 \citep{Gilli2013}. In summary, despite extensive research there is still considerable disagreement about the fraction of CT AGN and their contribution to the XRB particularly at high-redshift.  

In CT AGN, the majority ($>95\%$) of the hard X-ray (2 -- 10 keV) emission is obscured/scattered \citep{Risaliti1999,Matt2000}. The X-ray spectra, however, feature a prominent \feka\ emission line with large equivalent width EW $>$ 1 keV \citep[e.g.][]{Nandra1997,Reynolds2007,Vignali2002,Liu2016}, and the Compton hump, peaking at $\sim$20 -- 30 keV \citep{Krolik1999,Nandra2006}.  The spectral curvature method was developed by \cite{Kossinpreparation}, to identify nearby ($z<0.03$) CT AGN candidates in \swiftbat and \textit{NuSTAR} using the ($>10$ keV) spectral curvature.  The sensitivity of \textit{NuSTAR} is  $1\times 10^{-14}\;  \mathrm{erg\; cm^{-2}\; s^{-1}}$ in the 10 -- 30 keV range \citep{Harrison2013}, while \swiftbat has a sensitivity of $10^{-11}\;  \mathrm{erg\; cm^{-2}\; s^{-1}}$ in the deepest all sky maps \citep{Krimm2013}. Thus, both instruments select relatively bright sources compared to the faint high-redshift AGN detected by \chandrash.  The \textit{Chandra} Deep Field South (CDF-S), which is the deepest survey of the \textit{Chandra} X-ray observatory, has a flux limit of $5.5\times10^{-17}\; \mathrm{erg\; cm^{-2}\; s^{-1}}$ in the 2 - 10 keV energy range \citep{Xue2011}. Hence, it can detect much fainter sources than \textit{NuSTAR} such as high-redshift CT AGN.

In this article, we extend the Spectral Curvature (SC) method to high-redshift ($z>2$) AGN where the restframe Compton hump feature can be observed with \chandrash. In section 2 we describe our simulations to define the method for \textit{Chandra}, in section 3 we apply it to \chandra fields, and finally in section 4 we discuss implications.  Throughout this work, we adopt $\Omega_m$= 0.27, $\Omega_\Lambda$= 0.73, and $H_0$ = 71 km s$^{-1}$ Mpc$^{-1}$.  Errors are quoted at the 90\% confidence level unless otherwise specified.   

\section{The Spectral Curvature method}
\label{sec:SCmethod}

To estimate the likelihood of a X-ray source to be CT, the SC method uses the distinctive spectral shape of CT AGN at energies higher than 10 keV. In this work, we follow the technique used for low-redshift sources \citep[e.g.,][]{Kossinpreparation} where we model an unobscured source with a power-law of $\Gamma$ = 1.9, and a heavily CT source as an AGN with line-of-sight column densities of $\mathrm{N_H}=5 \times 10^{24}\; \mathrm{cm^{-2}}$ using the \mytorus spectral models from \cite{Yaqoob2012}. We choose the threshold at column density $\mathrm{N_H}=5 \times 10^{24}\; \mathrm{cm^{-2}}$ to be consistent with \citet{Kossinpreparation}. The SC equation is modeled so that an unobscured source has a SC value of zero, while an heavily CT AGN has a SC value of one.    

   \begin{figure*}
     \subfloat{%
       \includegraphics[width=0.52\textwidth]{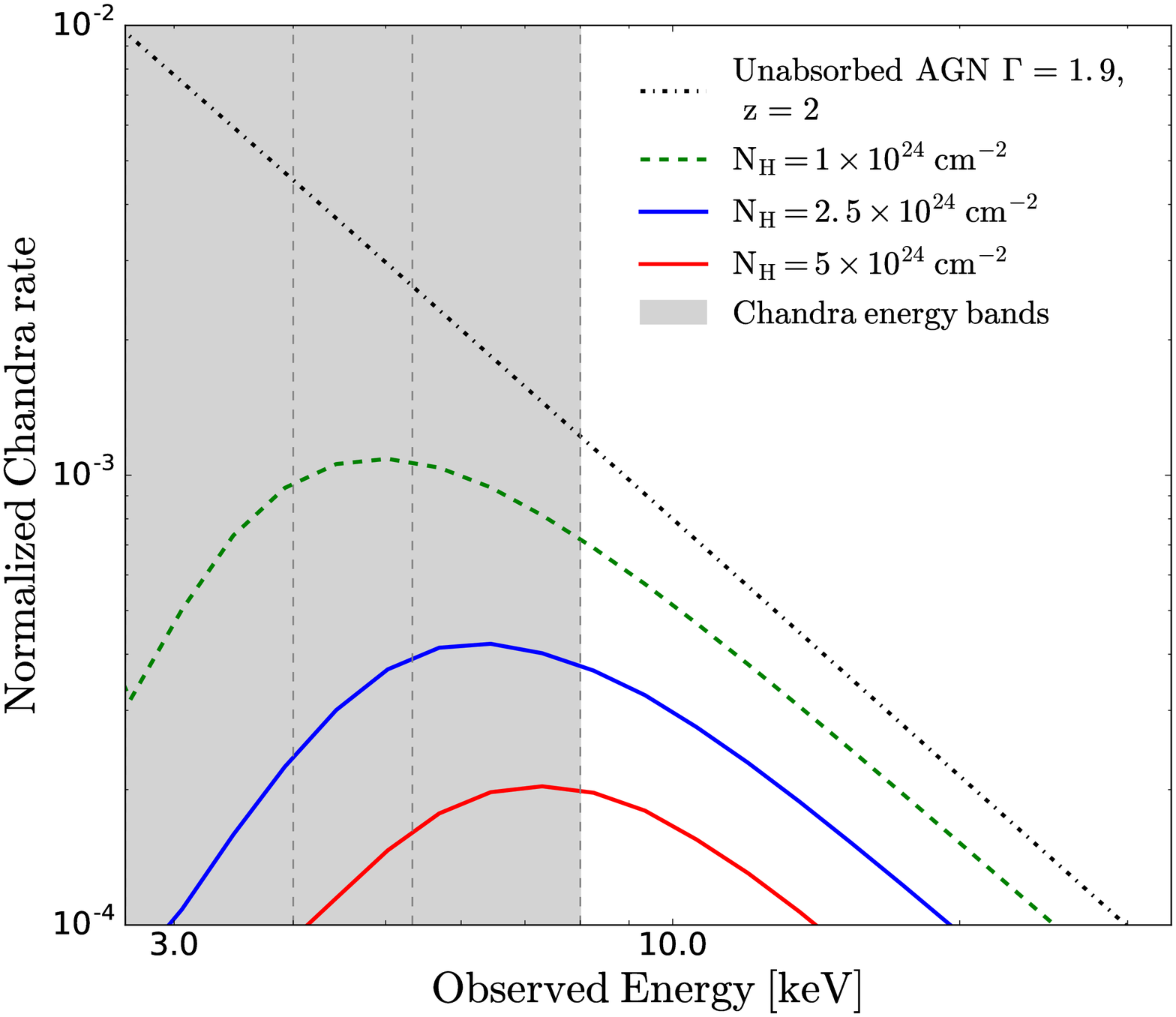}
     }
     \subfloat{%
       \includegraphics[width=0.5\textwidth]{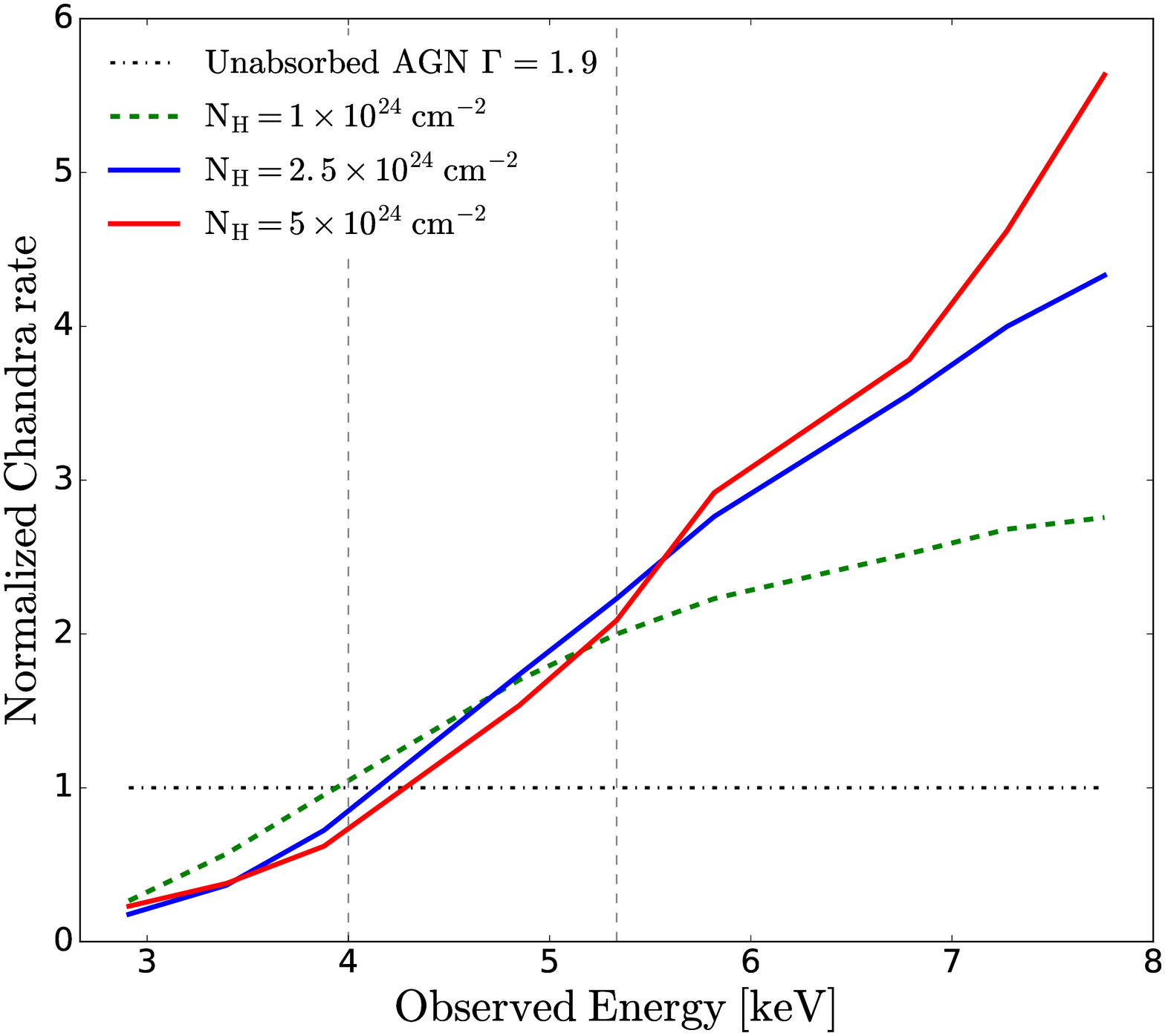}
     }
     \caption{\textit{Left}: CT AGN at redshift $z=2$ simulated using the \texttt{MYTorus} model compared to an unobscured power-law with $\Gamma=1.9$. As $\mathrm{N_H}$ increase from $\mathrm{10^{19}\, atoms^{-2}}$ to $\mathrm{5\times 10^{24}\, atoms^{-2}}$ the curvature of the observed radiation increases. At redshift $z>2$ the peak of the Compton hump is at rest frame energies smaller than 8 keV and can be observed with \chandrash. \textit{Right}: \textit{Chandra} number of counts for the same simulated sources at $z=2$ normalized by the total number of counts of the power-law source in the full energy band $[8-24]/(1+z)$ keV. The vertical dashed lines show the three energy ranges $[8-12]/(1+z)$, $[12-16]/(1+z)$ and $[16-24]/(1+z)$ keV. For energies above 4.5 keV (observed frame) the CT sources show an excess compared to the rate of an unobscured source. At energies below 4.5 keV, CT sources show a decrement.}
     \label{fig:Mikeplot}
   \end{figure*}
 As a first step, we simulated the Spectral Curvature (SC) of obscured AGN at high-redshift with the \texttt{XSPEC}  (version 12.9.0) {\tt fakeit} tool. Figure \ref{fig:Mikeplot}, left panel, shows how the SC measure increases with higher column density.  For simplicity, we assume $\mathrm{N_H} = 10^{24}\; \mathrm{cm^{-2}}$ as the lower limit of column density for CT AGN. The coefficients of the SC equation are defined using weighted and averaged counts of simulated unobscured and CT sources in three different energy ranges divided by the total counts in the entire range (8-24 keV restframe) (Figure \ref{fig:Mikeplot}, right panel). Finally, since we worked with observations of objects at redshift $z>2$, the corresponding energy ranges in the observed frame are $[8-12]/(1+z)$, $[12-16]/(1+z)$ and $[16-24]/(1+z)$ keV.

\subsection{The Spectral Curvature equation}
We first consider the importance of energy-dependent vignetting and point spread function degradation with off-axis angle. We tested the behavior of the SC equations for off-axis sources by simulating spectra of unobscured, obscured and CT AGN at constant redshift z = 2, exposure time (4 Ms) and intrinsic luminosity of $5\times 10^{44}\; \mathrm{erg\; s^{-1}}$, using response files corresponding to different off-axis positions. 
The response files at different off-axis angles are obtained using the \texttt{CIAO 4.9}  tools \texttt{mkacisrmf} and \texttt{mkarf}\footnote{See \url{cxc.harvard.edu/caldb/prop_plan/imaging/}.}. We averaged over 100 simulations to reduce the effect of Poisson noise. The coefficients of the spectral curvature equation show very little dependence on the off-axis position of the source in \textit{Chandra} (Figure \ref{fig:offaxis}). Nevertheless, we note that above 8 arcmin off-axis distance the large PSF significantly reduces sensitivity in \chandrash. On the other hand, the SC coefficients show a strong redshift dependency that can be corrected for using an additional redshift correction factor.

The SC equation for the \textit{Chandra} at redshift $z=2$ is given by:

\begin{equation}
\mathrm{{\textit{SC}_\textit{C}} (A, B, C)=-0.915\times A+ 0.281\times B+ 2.746\times C} \; ,
\end{equation}

where $\mathrm{A,\; B}$ and $\mathrm{C}$ are the normalized \textit{Chandra} count rates in the three energy ranges $[8-12]/(1+z)$, $[12-16]/(1+z)$ and $[16-24]/(1+z)$ keV, with $z=2$. The subscript \textit{C} indicates that we are referring to the \textit{Chandra} telescope.

The error on the SC values depends on the error on the counts, $\mathrm{\Delta A\;, \Delta B}$ and $\mathrm{\Delta C}$, which is given by the Poisson statistics. Thus, the error on the SC equation is:

\begin{equation}	
\mathrm{\Delta {\textit{SC}_\textit{C}}=\sqrt{(-0.915\times \Delta A)^2+(0.281\times \Delta B)^2+ (2.746\times \Delta C)^2}} \; ,
\end{equation}

We did not include the standard deviation on the calculated SC coefficient in the error propagation of the SC equation, since it is much smaller than the coefficient value itself and does not affect the total error much. Moreover, it is important to remember that A, B, and C are the counts in the three energy ranges \textit{normalized} for the counts F in the full energy band $[8-24]/(1+z)$ keV.  \\

\citealt{Kossinpreparation} showed that SC measurements are consistent for different telescopes. This means that we can apply the SC method to different satellites as, for example, the  \textit{XMM-Newton} and the future \textit{ATHENA} telescope.  The Wide Field Imager of the \textit{ATHENA} telescope will span the energy range from 0.1 to 15 keV. Finally, \textit{eROSITA} will scan the entire sky out to 10 keV.  We calculated the SC equation for \textit{ATHENA} and \textit{XMM-Newton} at $\mathrm{z = 1}$, since the two satellites can resolve the Compton hump starting from these redshifts because of their higher effective area at high energies. 
The SC equation for the different telescopes is given by:

\begin{equation}
\mathrm{{\textit{SC}_\textit{A}} (A, B, C)=-0.522\times A+ 0.251\times B+ 2.270\times C} \; ,
\end{equation}

\begin{equation}
\mathrm{{\textit{SC}_\textit{XMM}} (A, B, C)=-0.559\times A+ 0.424\times B+ 2.570\times C} \; ,
\end{equation}

\begin{equation}
\mathrm{{\textit{SC}_\textit{eROSITA}}(A, B, C)=-0.436\times A+ 0.407\times B+ 2.356\times C} \; ,
\end{equation}

\subsubsection{Redshift dependence}

After applying the method developed in \citet{Kossinpreparation} to the redshift interval from 2 to 5, the SC values and the thresholds between CT and non-CT sources depend significantly on the redshift.  We therefore add a redshift parameter to the SC equation, so that the new input variables are the normalized counts in the three energy ranges (A, B, and C) and include the change with redshift.\\

We develop an equation so that SC$>0.4$ is a consistent boundary for CT sources with redshift. We choose to normalize the threshold to a value of 0.4 to be consistent with \citet{Kossinpreparation}. We achieved this by calculating the SC values of simulated CT sources at different redshift. These values can be fitted with good approximation by a third degrees polynomial. We normalize the SC equation by this third degree polynomial to achieve the simplest model that provides a CT selection value with redshift close to a constant value. The CT threshold is still slightly redshift dependent since the third degree polynomial only approximates the curve that describes the SC values of CT sources. The redshift correction factor is then:

\begin{equation}
\mathrm{\textit{CT}_{\textit{C}}(z) = -0.02\times z^3+0.29  \times z^2 +3.00 \times z -3.35 }\; ,
\end{equation}

\begin{equation}
\mathrm{\textit{CT}_{\textit{A}}(z) = 0.03\times z^3-0.41  \times z^2 +1.98 \times z -0.78 }\; ,
\end{equation}

\begin{equation}
\mathrm{\textit{CT}_{\textit{XMM}}(z) = 0.03\times z^3-0.40  \times z^2 +1.37 \times z -0.64 }\; 
\end{equation}

\begin{equation}
\mathrm{\textit{CT}_{\textit{eROSITA}}(z) = 0.04\times z^3-0.54 \times z^2 +2.07 \times z -1.40 }\; 
\end{equation}

The new SC equation has the form
\begin{equation}
\mathrm{{\overline{\textit{SC}}}_{I}}(A,B,C,z) = \frac{{SC_{I}}(A,B,C)}{{CT_{I}}(z)},
\end{equation}

where \textit{I} is $\{$\textit{C}, \textit{A}, \textit{XMM}, \textit{eROSITA}$\}$.

\begin{center}
\begin{figure}
\centering
\includegraphics[scale=0.34]{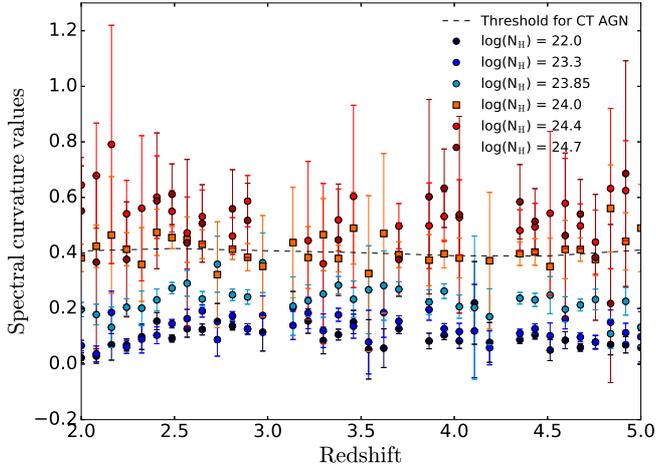}
\caption{Normalized SC values of simulated sources versus the redshift. The SC method selects sources with column densities above $10^{24}$ cm$^{-2}$ as CT AGN. Thus, the SC method successfully distinguish CT sources from merely obscured AGN.}
\label{fig:red}
\end{figure}
\end{center}

\begin{center}
\begin{figure}
\centering
\includegraphics[scale=0.35]{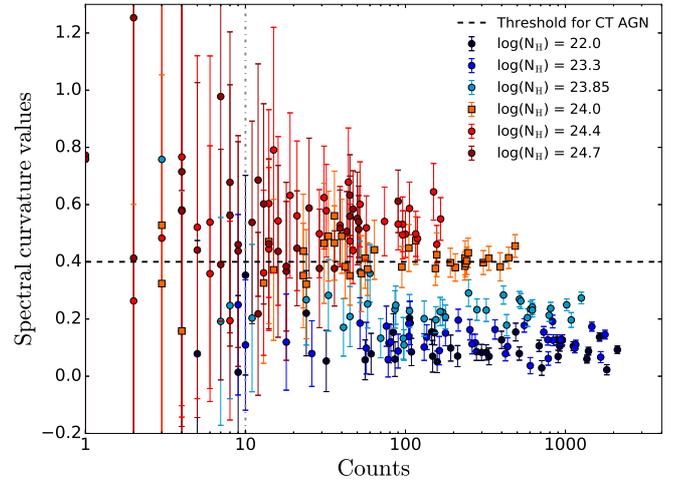}
\caption{Evolution of the SC values as function of column density and counts in the full energy range. Note that for smaller number of counts the error bars become larger. For less than 10 counts in the full energy range, the SC value is unreliable. The spectra of non-CT and CT sources are simulated using the \texttt{fakeit} tool of \texttt{XSPEC} by constant exposure time of 4 Ms, this determines the upper limit in the number of counts of the spectra.}
\label{fig:counts}
\end{figure}
\end{center}

We tested the SC method on a sample of simulated X-ray spectra with different column densities and luminosities. The integration time for the simulation is set to 4 Ms, this determines a limit on the maximum number of counts obtained. From Figure \ref{fig:red} and Figure \ref{fig:counts} we observe that the method successfully distinguishes between simulated sources with column densities below $\mathrm{N_H = 10^{24}\; \, cm^{-2}}$ and CT sources. Moreover, from Figure \ref{fig:counts} we can estimate where the SC method is less reliable for sources with very few counts due to the large error bars. We note that between 10 and 70 counts the SC method presents large uncertainties that could make uncertain the classification for single sources, however, the population can be studied in aggregate. Moreover, the method is less sensitive to sources with column densities exceeding $\mathrm{N_H = 10^{25}\; \, cm^{-2}}$, since at these column densities the Compton hump intensity is reduced by Compton scattering. Thus, the SC method is better suitable for transmission-dominated ($\mathrm{N_H< 10^{25}\; \, cm^{-2}}$) CT AGN.

\begin{center}
\begin{figure}\label{fig:offaxis}
\centering
\includegraphics[scale=0.34]{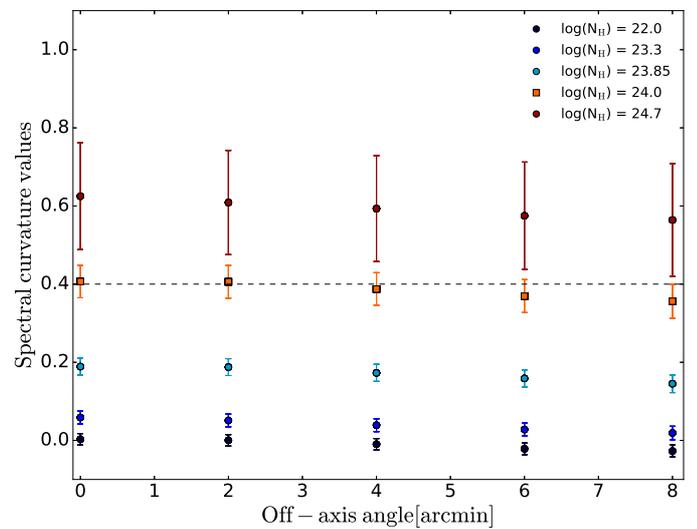}
\caption{Spectral curvature values as function of the off-axis position for simulated AGN with different column densities $\mathrm{N_H}$, redshift $z=2$ and with intrinsic luminosity of $5\times 10^{44}\; \mathrm{erg\; s^{-1}}$. The deviations in the SC values due to the off-axis position are small compared to the error bars.}
\label{fig:offaxis}
\end{figure}
\end{center}

\section{Sample and data analysis}
\label{sec:sample_data}

We applied the SC method to deep \textit{Chandra} observations. Thanks to the high sensitivity of \textit{Chandra}, we can find CT candidates even at redshift higher than 2. The deepest \textit{Chandra} surveys are the CDF-S and the COSMOS legacy survey (Figure \ref{fig:COSMOSsample}).

\subsection{CDF-S} \label{SamplesCDFS}
The \textit{Chandra} Deep Field South (CDF-S) has an on axis flux limits reaches 3.2$\times$10$^{-17}$, 9.1$\times$10$^{-18}$ and 5.5$\times$10$^{-17}$ erg cm$^{-2}$ s$^{-1}$ in the energy ranges 0.5 -- 8 keV, 0.5 -- 2 keV, and 2 -- 8 keV respectively \citep{Xue2011}  (Figure \ref{fig:COSMOSsample}). For this catalog, the reduced spectra have not been made public. Thus, we applied the SC method on the 4 Ms merged event file\footnote{The event file can be found on the CXC homepage (\url{http://cxc.harvard.edu/cda/Contrib/CDFS.html})}. The coordinates and redshifts that we used, can be found in the catalog of \citet{Xue2011}.

We excluded from the analysis the sources with angular distance greater than 8.7 arcmin from the image center (Figure \ref{fig:offaxis}) because of their significantly reduced sensitivity and exposure time. 
We extracted the net number of counts and the error on it using the \texttt{CIAO} tool \texttt{dmextract}. We extracted the net counts in the three energy ranges $[8-12]/(1+z)$, $[12-16]/(1+z)$ and $[16-24]/(1+z)$ keV. The error on the net counts are calculated directly with \texttt{dmextract} using the Gehrels statistic \citep{Gehrels1986}. The flux limit for the CDF-S is calculated for unobscured sources. Hence, the survey may miss sources with very high level of obscuration that fall below the detection limit, for example the reflection-dominated CT AGN ($\mathrm{N_H > 1\times 10^{25}\, cm^{-2}}$).

\subsection{COSMOS}
\begin{center}
\begin{figure}
\centering
\includegraphics[scale=0.35]{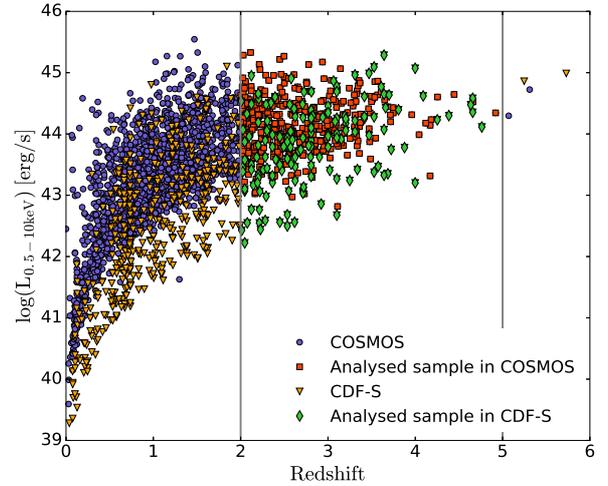}
\caption{Luminosity in the 0.5 -- 10 keV range compared with the redshift. The black dashed line shows the flux limit in the full energy range. The squared points indicate the sources taken into account in this work. Data from \citet{Marchesi2016}.}
\label{fig:COSMOSsample}
\end{figure}
\end{center}

The \textit{Chandra} COSMOS Legacy Survey covers 2.2sq deg of the COSMOS field to a flux limit of 2.2$\times$ 10$^{-16}$, 1.5$\times$ 10$^{-15}$ and 8.9$\times$ 10$^{-16}$ erg cm$^{-2}$  s$^{-1}$ in the 0.5 -- 2, 2 -- 10 and 0.5 -- 10 keV bands, respectively \citep{Civano2016}  (Figure \ref{fig:COSMOSsample}). The depth of the flux and the relatively large area of the COSMOS-Legacy survey are going to remain unrivaled until the advent of \textit{ATHENA} \citep{Civano2016}.


We used the X-ray spectra of the sources in the COSMOS-legacy survey from \citet{Civano2016}. For the purposes of our analysis, for each source in the \textit{Chandra} COSMOS-Legacy sample we used  \texttt{XSPEC} (version 12.9.0) to estimate the number of counts in the energy intervals $[8-12]/(1+z)$, $[12-16]/(1+z)$ and $[16-24]/(1+z)$ keV. The number of counts in the energy range is calculated by multiplying the count rate (obtained by calling the attribute \texttt{rate} of \texttt{xspec.Spectrum}) with the exposure time.
 To determine the errors on the number of counts, we applied Gehrels statistic \citep{Gehrels1986}.  The exposure times of our sample range from 40 to 250 ks.
We did not exclude sources at largest off axis angles. The flux limit for the COSMOS survey is calculated for unobscured source. Therefore, highly obscured sources are likely to be missed from the survey.

\section{Results}
\label{sec:Results}

\subsection{CDF-S}

\begin{center}
\begin{figure}
\centering
\includegraphics[scale=0.3]{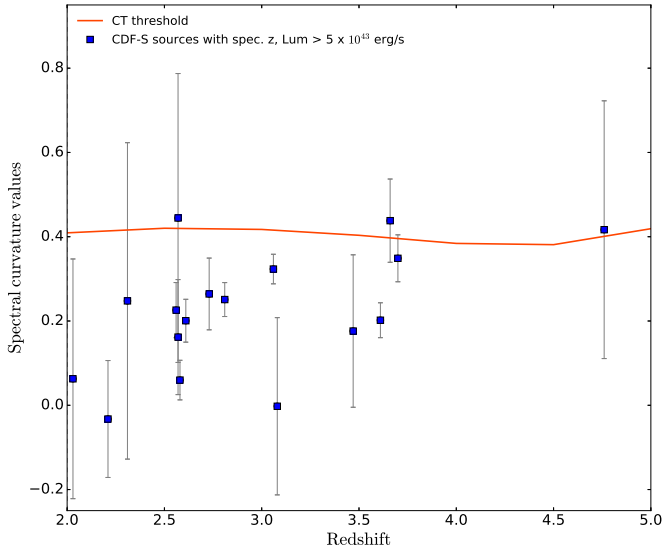}
\caption{CDF-S spectral curvature values for sources with spectroscopic redshift. The red line shows the threshold between non-CT and CT AGN. The sources we show here are enclosed in a region with radius of 8.7 arcmin from the field center to avoid the sources with extremely large PSF. To take into account only the SC values with smaller error bars, we apply the SC method only on sources with luminosity higher than $5 \times 10^{43}$ $\mathrm{erg\,s^{-1}}$. The source at redshift $z=4.67$ has been proposed as a CT AGN by \citet{Gilli2011}. The SC method selects this source as a CT candidate but still with a large error bar.}
\label{fig:CDFS_SC}
\end{figure}
\end{center}

By applying the SC method on 17 sources with luminosity higher than $5\times 10^{43}$ $\mathrm{erg\,s^{-1}}$ and with spectroscopic redshift in the 4 Ms CDF-S (see Table \ref{table:CDFS}), we obtained three CT candidates (the three blue dots above the CT threshold line in Figure \ref{fig:CDFS_SC}). Of these, only the one at redshift 3.66 has net number of counts higher than 100 (see Figure \ref{fig:CDFS_SC}, middle source above the threshold).  The source above the CT threshold at redshift 4.67 has  coordinates RA = 3:32:29.27, Dec = -27:56:19.8 (XID403) and has been proposed as a CT candidate in \citet{Gilli2011}.  The SC value could be verified by applying the SC method to the coming 7 Ms CDF-S survey which will have tighter limits and smaller uncertainties. 

Constraining our analysis to sources with spectroscopic redshift, the fraction of CT AGN selected in the CDF-S is 17$^{+18.6}_{-11.0}\%$ (3/17 sources with spectroscopic redshift) assuming binomial statistics with 90$\%$ of confidence. The value we obtain is similar to what was found by \citet{Kossinpreparation}.  However, the sample analyzed in the CDF-S is small. To have more reliable constraints on the CT AGN populations we will focus on the larger sample obtained from the COSMOS-legacy survey.

\subsection{COSMOS}
\label{res_COSMOS}
\begin{figure}
     \subfloat{%
       \includegraphics[width=0.5\textwidth]{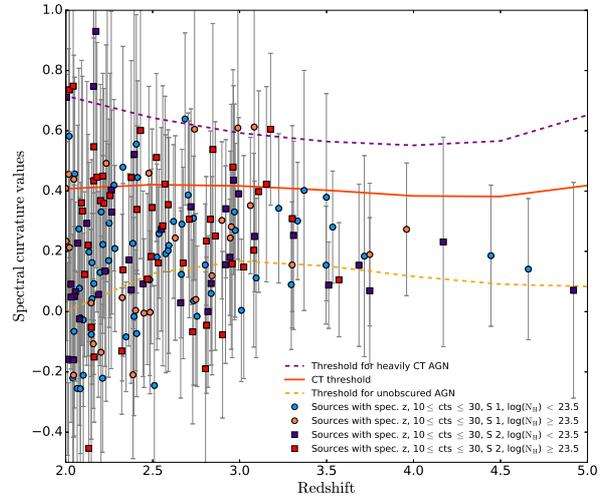}
     }
     \hfill
     \subfloat{%
       \includegraphics[width=0.5\textwidth]{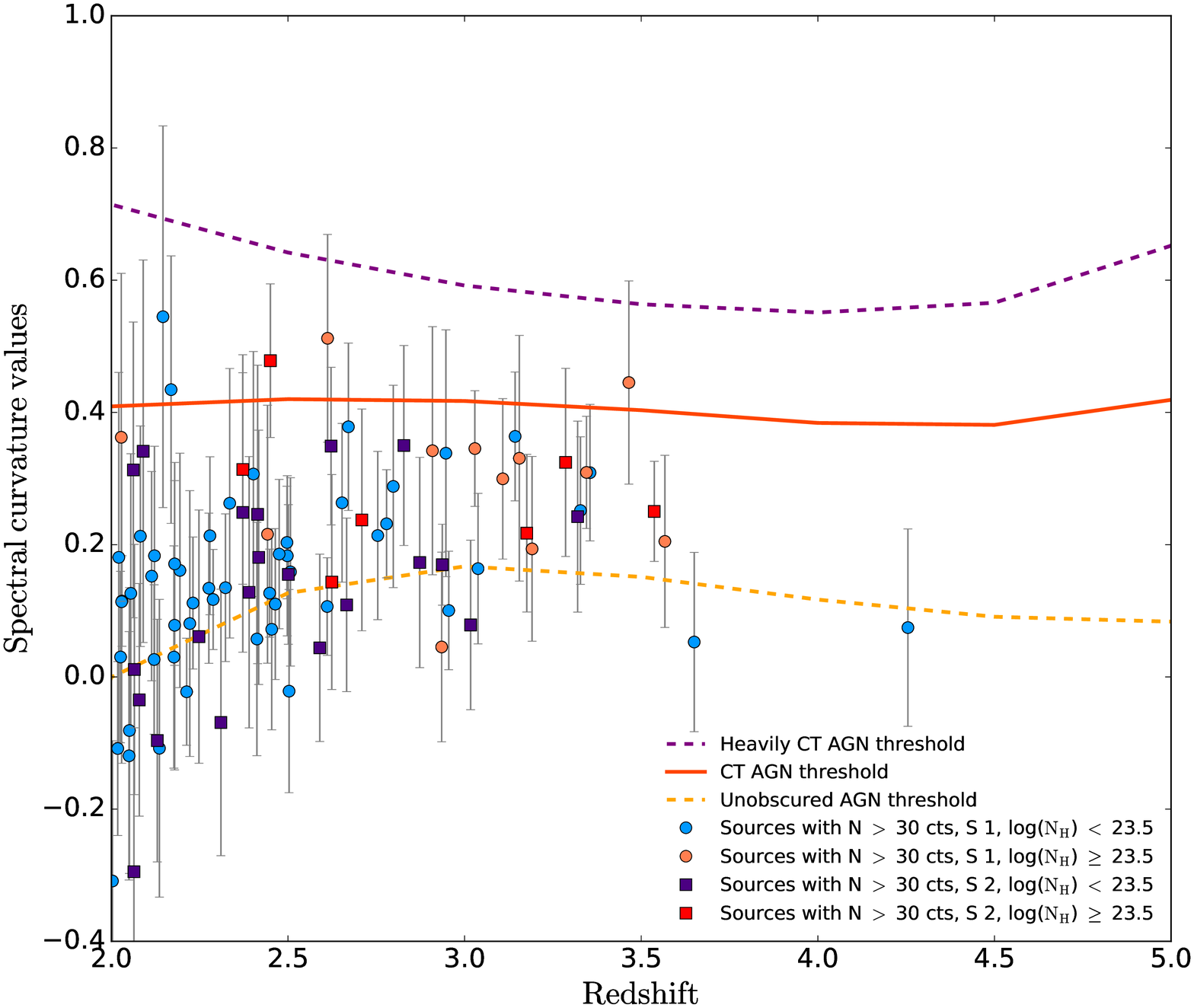}
     }
     \caption{\textit{Top panel}: The fraction of sources with $\mathrm{N_H}<23.5$ selected as CT candidates is 8.6$^{+4.3}_{-3.2}\%$. Of the sources (120) cataloged with column density $\mathrm{log(N_H)\geq 23.5}$ 54, are cataloged as Seyfert 1. The fraction of CT candidates selected for this $\mathrm{N_H}$ range are $22.5^{+6.7}_{-5.7}\%$. \textit{Bottom panel}: Same as above but showing only the detected sources with more than 30 counts. 80$\%$ of the CT candidates (4/5 sources) agree with the spectral measurements.}
     \label{fig:COSMOS_SCnH}
   \end{figure}

We calculated the SC values for the sources in the COSMOS-legacy survey between redshift 2 and 5 (Figure \ref{fig:COSMOS_SCnH}). The redshift and the column densities of the COSMOS sources can be found in the \citet{Marchesi2016} catalog. The $\mathrm{N_H}$ therein are calculated from HR ratios and redshifts. 
 In total we applied the method to 272 sources (see Table \ref{table:COSMOS}), 158 cataloged as Seyfert 1 (i.e. unobscured AGN showing both broad and narrow optical emission lines) and 68 cataloged as Seyfert 2 (i.e. obscured AGN showing only narrow optical emission lines).

We found that 14.5$\%$ (40/272) sources are selected as CT AGN. 
The SC method selects no CT candidate at redshift $z>3.5$, primarily due to the much smaller number of sources in the survey and their faintness. If we restrict the luminosity to $\mathrm{L_X>10^{44}\, erg\, s^{-1}}$ to avoid biases due to the flux limit of the COSMOS survey we find that the fraction of CT AGN is 8.9$\%$ (13/145). We chose to apply this luminosity cut, since we are comparing the obtained CT fraction in different redshift bins over a specific luminosity range so that it can be compared to other published studies (e.g. \citet{Ricci2015}) and because of the low statistical significance of the spectral curvature for sources just above the detection limit.  The focus on higher luminosity AGN in this paper will likely exclude some number of absorbed AGN because of the well-known anti-correlation between fraction of absorbed AGN and luminosity (e.g. \citet{Hasinger2008}). Considering the total sample of CT candidates (without luminosity cuts), 18/40 (45$\%$) are described as Seyfert 1 in the catalog. This means that 11.4$\%$ of sources that are considered unobscured in the optical is selected as CT AGN candidates.  On the other hand, 22 sources are selected from the 68 cataloged as Seyfert 2. This means that the 32.4$\%$ of the Seyfert 2 is selected as CT. We also have to consider the possibility that the classification of Seyfert 1 and Seyfert 2 in the \citet{Marchesi2016} catalog might have some uncertainties. Additionally, the definitions of Seyfert 1 and Seyfert 2 are based on optical spectra and X-ray (unobscured vs. obscured) schemes of classification do not always agree \citep[e.g.,][]{Burtscher2016}. To explore this possibility, we examined the SC values of the COSMOS sample for column densities $\mathrm{log(N_H)}<23.5$ cm$^{-2}$ and $\mathrm{log(N_H)}\geq 23.5$ cm$^{-2}$ (Figure \ref{fig:COSMOS_SCnH}, bottom panel). We found that only $8.6^{+4.3}_{-3.2}\%$ of the sources with $\mathrm{log(N_H)}<23.5$ cm$^{-2}$ (in total 152) are selected as CT candidates, while we select as CT candidates $22.5^{+6.7}_{-5.7}\%$ of the sources with $\mathrm{log(N_H)}\geq 23.5$ cm$^{-2}$ (in total 120). This means that the SC method typically agrees with CT AGN candidates sources with high values of $\mathrm{N_H}$.


We also compare our results with the $\mathrm{N_H}$ obtained from spectral fitting by \cite{Marchesi2016a}. The only source with $\mathrm{N_H>10^{24}\, cm^{-2}}$ reported in \cite{Marchesi2016a} is Cid$\_$747. Its SC value is $\mathrm{0.24\pm0.17}$ and thus the source is not selected as CT candidate by the SC method. Larger samples of CT AGN from X-ray spectral fitting would be useful for further comparison.\\

The mean value of spectral curvature for Seyfert 1 is 0.16$\pm$0.02, while for the Seyfert 2 we have a mean value of 0.26$\pm$0.03. This is a promising result, since Seyfert 2 defines sources obscured in the optical wavelengths and thus we expect to find all CT candidates in the Seyfert 2 population. \\

However, the number of selected Seyfert 1 is high and has to be investigated whether this is statistical noise.   To test this we assumed that all the Seyfert 1 sources should be completely unobscured, i.e. a pure power-law, and their SC values should be zero. Then we randomly added noise consistent with the expected uncertainty. We repeated this 100 times in a bootstrap process to estimate the error. The average number of sources selected as CT AGN is 11.6$\%$ with standard deviation of 3.1$\%$, which is consistent with the fraction of selected Seyfert 1 suggesting this population is consistent with the false positive expected from statistical noise.\\

We also predict the false positive and negative rates of the SC method by inferring them from simulations. For simplicity, we assume a flat \NH\ distribution of sources with equal numbers at every column density between $10^{21}$ \nhunit\ and $5\times10^{24}$ \nhunit.  Sources between $10^{21}$ \nhunit\ and $10^{24}$ \nhunit\ can contribute to false positives and sources with $\mathrm{N_H}$ between $10^{24}$ \nhunit\ and $5\times10^{24}$\nhunit\ can be missed false negatives because of statistical noise. At the exposure times in COSMOS, we found that the rate of false positives is between $9\%$ to $16\%$ for the shortest and longest exposures which is consistent with our previous false positive rate measurement. The rate of false negatives varies between $23\%$ and $44\%$ between the shortest and longest exposure suggesting that a significant fraction of transmission dominated CT AGN will be missed.\\

We also have to consider the possibility that the classification of Seyfert 1 and Seyfert 2 in the \citet{Marchesi2016} catalog might have some uncertainties. Additionally, the definitions of Seyfert 1 and Seyfert 2 are based on optical spectra and X-ray (unobscured vs. obscured) schemes of classification do not always correspond \citep{Burtscher2016}. To explore this possibility, we examined the SC values of the COSMOS sample for column densities $\mathrm{log(N_H)}<23.5$ cm$^{-2}$ and $\mathrm{log(N_H)}\geq 23.5$ cm$^{-2}$ (Figure \ref{fig:COSMOS_SCnH}). We found that only $8.6^{+4.3}_{-3.2}\%$ of the sources with $\mathrm{log(N_H)}<23.5$ cm$^{-2}$ (in total 152) are selected as CT candidates, while we select as CT candidates  $22.5^{+6.7}_{-5.7}\%$ of the sources with $\mathrm{log(N_H)}\geq 23.5$ cm$^{-2}$ (in total 120). This means that the SC method effectively selects as CT AGN candidates sources with high values of $\mathrm{N_H}$. Of the sources with $\mathrm{log(N_H)}<23.5$ cm$^{-2}$, 104 are cataloged as Seyfert 1 while 48 are tagged as Seyfert 2. In the $\mathrm{log(N_H)}\geq 23.5$ cm$^{-2}$ regime, 66 source are considered as Seyfert 2 and 54 as Seyfert 1. However, since their line-of-sight column density is quite high, they can not be considered to be unobscured sources in the X-ray. In the high $\mathrm{N_H}$ case, $20.4\%^{+10.0\%}_{-7.8\%}$ Seyfert 1 and $24.2^{+9.4}_{-7.7}\%$ Seyfert 2 are selected as CT candidates.\\

Another possible explanation for the fraction of Seyfert 1 selected as CT candidates is that at these redshifts the reflection component of their X-ray spectra enters in the energy range we examine with the SC method. However, \citep[e.g.,][]{Kossinpreparation} tested a larger range of torus models and found that these sources would still be well below the CT limit.\\

Another issue is that the observed luminosity of faint CT sources will be below the flux limit of the survey.  We therefore perform simulations to correct for highly obscured sources missed with \chandrash.  We calculate the ratio of intrinsic to observed luminosity as a function of redshift and column density in the rest frame energy band from 8 -- 24 keV by simulating sources with different $\mathrm{N_H}$. Using this value, we can calculate which fraction of sources we are not able to detect with \textit{Chandra} in different redshift bins and for different $\mathrm{N_H}$. \\

To estimate the fraction of faint undetected sources, we randomly draw the $\mathrm{N_H}$ of the simulated sources from two different $\mathrm{N_H}$ distributions at $z=2$: a linear distribution and the observed $\mathrm{N_H}$ distribution proposed by \citet{Ricci2015} and we calculate the fraction of sources too faint for \textit{Chandra} to observe if the $\mathrm{N_H}$ is the one assumed. We obtained this fraction by simulating a population of unobscured sources using the \texttt{fakeit} tool of \texttt{XSPEC} and by comparing how many of these sources are below \textit{Chandra} sensitivity if we apply the randomly draw $\mathrm{N_H}$. The luminosities and redshift of the unobscured simulated sources are comparable with those of the sources in the COSMOS sample. The integration time of the simulations is held constant to 4 Ms consistent with the survey. We repeat this calculation 1000 times, each time drawing a new random sample from the parent $\mathrm{N_H}$ distribution. 
Since we have a fraction of CT AGN equal to zero above z=3.5, we constrain this analysis to z = 2--3.5. The percentage of non detected sources in the redshift range z = 2--3.5 is 42.6$\%$ for the linear distribution and 44$\%$ for the $\mathrm{N_H}$ in \citet{Ricci2015}.  While the correction factor would be different in the cases of a $\mathrm{N_H}$ distribution centered on very low or very high column density values, observational and empirical estimations from \citet{Ricci2015} and \citet{Ueda2014} make this scenario unlikely.\\

\begin{figure*}
\centering
\includegraphics[scale=0.5]{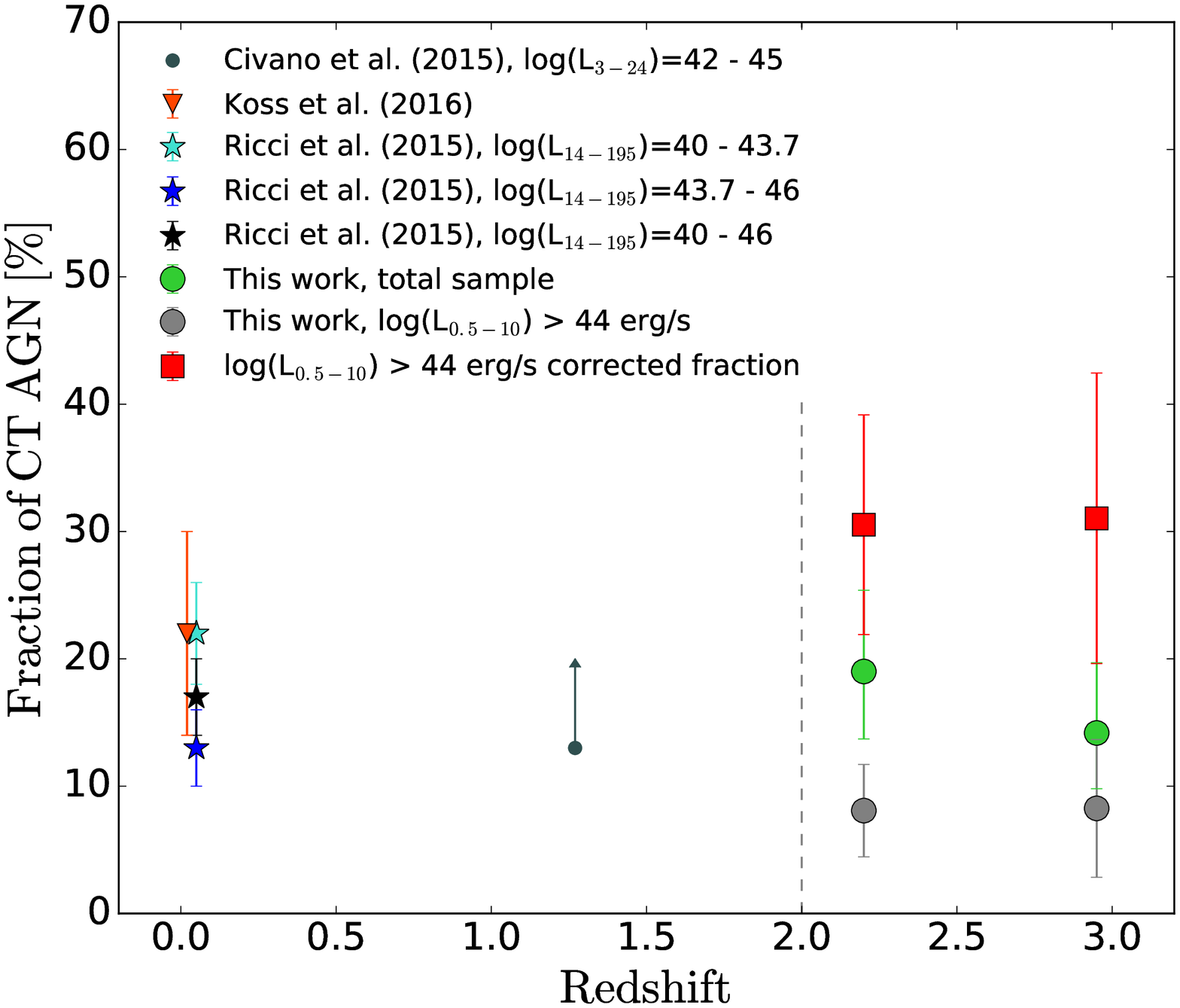}
\caption{Fraction of CT AGN in sample obtained from the COSMOS survey (in green) compared with the CT fraction obtained by \protect\cite{Kossinpreparation} from the \textit{BAT} catalog (orange), to the CT fraction found by \protect\cite{Ricci2015} in the nearby Universe and to the fraction obtained by \protect\cite{Civano2015} in the redshift range z=0.04--2.5 (dark gray). Using simulations, we calculated the fraction of CT AGN that we are not able to detect due to \textit{Chandra} flux limit. The fraction of CT AGN from the COSMOS catalog corrected for the fraction of CT AGN we are not able to detect (in red) shows a constant behavior.}
\label{fig:fraction}
\end{figure*}

We restrict our sample to the redshift range from 2 to 3.5 and the luminosity range of $\mathrm{L_X>10^{44}\,erg \, s^{-1}}$  to estimate the intrinsic fraction of CT AGN due to the difficulty correcting for flux sensitivity limits.  This leads to an observed CT fraction of 8.9$\%$ (13/145).  Assuming the observed column density distribution of \citet{Ricci2015} and the flux sensitivity limits of the COSMOS legacy survey, the fractions of CT sources falling below the flux limit are $85\%$ and $87\%$, respectively, in the redshift bins from 2 to 2.7 and 2.7 to 3.5 (see Figure \ref{fig:fraction}). The fractions of non-CT sources that we do not detect in the same redshift bins are $28\%$ and 29$\%$.  After applying the correction, we find that the fraction of CT AGN in COSMOS is $\sim 32\pm11\%$.\\

\section{Summary and Conclusions}
\label{sec:Summary_Conclusion}

We extended the Spectral Curvature method, developed by \citet{Kossinpreparation}, to \textit{Chandra} observations at redshifts between 2 and 5. We summarize in the following our main findings:

\begin{itemize}
\item The SC method can be applied to high-redshift AGN observations. The redshift dependence can be corrected by adding a redshift parameter in the SC equation. The method successfully selects simulated CT sources from merely obscured ones.

\item We applied the SC method to the CDF-S survey. The SC method selects three sources as CT candidates. One of these is the proposed CT AGN from \citet{Gilli2011} at redshift $z=4.67$. The fraction of CT AGN we selected from the sources with spectroscopic redshift is 17$^{+19}_{-11}\%$.

\item We applied the method also to the COSMOS-legacy survey, constraining our analysis to sources at redshifts between 2 and 5 with more than 10 counts. In total, the method selected 40 from 272 sources as CT candidates (14.5$\%$). After correcting for biases due to the redshift and accounting for the faint sources that  \textit{Chandra} is not able to detect, we obtain a CT fraction of $\sim 32\pm10\%$ which is a value similar to the one found in \citet{Buchner2015}.

\item We find that the fraction of CT AGN does not show redshift evolution, which is comparable to the result found by \citet{Buchner2015} in the luminosity range $\mathrm{L_{2-10keV}=10^{43.2-46}\, erg\, s^{-1}}$. However, the fraction that we obtain is similar to the one found by \citet{Ricci2015} in the nearby Universe and by \cite{Civano2015} at redshift z=0.04--2.5 and much lower than the one obtained by \citet{Buchner2015}. This could be explained by the larger luminosity range analyzed in \citet{Buchner2015}.

\end{itemize}

Our measured CT fraction from COSMOS is somewhat higher though in agreement within error of the value of 22$\%$ found by \citet{Kossinpreparation}.  The mean luminosity of our sample is $\sim 10^{44}$ $\mathrm{erg\,s^{-1}}$, while the mean luminosity of the \textit{BAT} sources at redshift $z<0.03$ is $\sim 5\times 10^{42}$ $\mathrm{erg\,s^{-1}}$.  The fraction that we obtain is similar to what has been found by \citet{Ricci2015} in the lowest luminosity bin $\mathrm{log(L_{14-195})}=$40.0--43.7 $\mathrm{erg\,s^{-1}}$.  \\

Moreover, the SC method is insensitive to CT AGN of very higher column densities, e.g. $10^{25-26}$ cm$^{-2}$ which would not be detected in the X-rays. This means that we have to treat the obtained CT fraction as a lower limit.  Indeed, the obtained fraction of CT sources is lower than predicted by the models from \citet{Gilli2007} and \citet{Treister2009}.  Another issue is that the accretion rates of CT AGN may be much higher than their less obscured counterparts \cite[e.g.,][]{Kossinpreparation} and thus even a small fraction may be important for overall black hole growth.

As a further step, the SC method could be extended to other data samples. For example, the serendipitous \textit{Chandra} Multiwavelength Project (ChaMP) contains a number of promising high-redshift quasars that could satisfy the requirement needed to apply the method.  In the coming months, \textit{Chandra} will perform an observation of the CDF-S totaling 3 Ms which will complete the present survey. The deeper exposures in the 7 Ms catalog will allow tighter constraints on the fraction of CT AGN at high-redshift.

\begin{table*}
\caption{SC values for the analyzed sample in the CDF-S.}
\begin{center}
\begin{tabular}{cccccccc}
\hline
\hline
\multicolumn{1}{l}{XID$^{a}$} & z$^{b}$ & Soft cts$^{c}$ & Mid cts$^{d}$ & Hard cts$^{e}$ & Total cts$^{f}$ & SC$^{g}$ & CT candidate\\ \hline
710 & 2.03 & 45$\pm$10 & 17$\pm$10 & 16$\pm$10 & 77$\pm$17 & 0.06$\pm$0.29 &\\ 
369 & 2.21 & 101$\pm$17 & 54$\pm$13 & 25$\pm$13 & 180$\pm$25 & -0.03$\pm$0.14 & \\ 
20 & 2.31 & 21$\pm$9 & 7$\pm$7 & 13$\pm$8 & 42$\pm$14 & 0.25$\pm$0.38 &  \\ 
188 & 2.56 & 127$\pm$14 & 90$\pm$13 & 88$\pm$13 & 306$\pm$23 & 0.23$\pm$0.07  &\\ 
93 & 2.57 & 44$\pm$10 & 31$\pm$9 & 25$\pm$9 & 100$\pm$16 & 0.16$\pm$0.14 & \\ 
294 & 2.57 & 17$\pm$8 & 4$\pm$7 & 20$\pm$8 & 41$\pm$13 & 0.44$\pm$0.34 & yes\\ 
687 & 2.58 & 268$\pm$22 & 167$\pm$18 & 98$\pm$16 & 533$\pm$33 & 0.06$\pm$0.05  &\\ 
137 & 2.61 & 147$\pm$14 & 109$\pm$12 & 95$\pm$12 & 351$\pm$22 & 0.20$\pm$0.05  &\\ 
86 & 2.73 & 89$\pm$15 & 55$\pm$12 & 74$\pm$13 & 218$\pm$23 & 0.26$\pm$0.09  &\\ 
149 & 2.81 & 197$\pm$17 & 172$\pm$15 & 168$\pm$16 & 537$\pm$28 & 0.25$\pm$0.04 & \\ 
546 & 3.06 & 201$\pm$16 & 185$\pm$15 & 247$\pm$17 & 633$\pm$28 & 0.32$\pm$0.04 &  \\ 
674 & 3.08 & 26$\pm$8 & 11$\pm$7 & 7$\pm$8 & 45$\pm$13 & -0.00$\pm$0.21 &  \\ 
588 & 3.47 & 27$\pm$8 & 5$\pm$6 & 18$\pm$8 & 50$\pm$13 & 0.18$\pm$0.18 & \\ 
563 & 3.61 & 162$\pm$15 & 89$\pm$12 & 127$\pm$14 & 378$\pm$23 & 0.20$\pm$0.04 & \\ 
262 & 3.66 & 24$\pm$7& 33$\pm$8 & 60$\pm$10 & 117$\pm$14 & 0.44$\pm$0.10 & yes \\ 
412 & 3.7 & 65$\pm$10 & 74$\pm$10 & 108$\pm$12 & 247$\pm$18 & 0.35$\pm$0.06 & \\ 
403 & 4.76 & 9$\pm$5 & 4$\pm$4 & 12$\pm$5 & 24$\pm$8 & 0.42$\pm$0.31  & yes\\ \hline \hline
\end{tabular}
\end{center}
\label{table:CDFS}
\flushleft \footnotesize{$^{a}$Identification number of the source in the 4 Ms CDF-S \citep{Xue2011}.}\\
\footnotesize{$^{b}$Spectroscopic redshift from \cite{Xue2011}.}\\
\footnotesize{$^{c}$Number of counts in the soft energy range $\mathrm{[8-12]/(1+z)}$ keV.}\\
\footnotesize{$^{d}$Number of counts in the mid energy range $\mathrm{[12-16]/(1+z)}$ keV.}\\
\footnotesize{$^{e}$Number of counts in the hard energy range $\mathrm{[16-24]/(1+z)}$ keV.}\\
\footnotesize{$^{f}$Number of counts in the total energy range $\mathrm{[8-24]/(1+z)}$ keV.}\\
\footnotesize{$^{g}$Measured Spectral Curvature values.}
\end{table*}

\begin{table*}
\caption{SC values for the analyzed sample in the COSMOS legacy survey. This table is available in its entirety in a machine-readable form in the online journal. A portion is shown here for guidance regarding its form and content.}
\begin{center}
\begin{tabular}{cccccccccc}
\hline
\hline
\multicolumn{1}{c}{ID$^{a}$} & z$^{b}$& Soft cts$^{c}$ & Mid cts$^{d}$& Hard cts$^{e}$ & Total cts$^{f}$ & SC$^{g}$& CT candidate & $\mathrm{N_H}^{h}$ [$\mathrm{10^{22}}$ atoms/$\mathrm{cm^2}$] &$\mathrm{N_H}^{i}$ [$\mathrm{10^{22}}$ atoms/$\mathrm{cm^2}$] \\ \hline
Lid$\_$471 & 2.0 & 7$\pm$4 & 8$\pm$4 & 6$\pm$3 & 23$\pm$6 & 0.41$\pm$0.38 & yes &18.8 &16.10 \\ 
Lid$\_$1026 & 2.00 & 10$\pm$4 & 9$\pm$4 & 5$\pm$3 & 25$\pm$6 & 0.23$\pm$0.33 & & 25.3 & 2.67 \\ 
Lid$\_$249 & 2.00 & 24$\pm$6 & 6$\pm$3 & 2$\pm$3 & 33$\pm$6 & -0.31$\pm$0.21& & 0 & 4.22 \\ 
Cid$\_$545 & 2.01 & 14$\pm$5 & 6$\pm$3 & 2$\pm$3 & 23$\pm$6 & -0.16$\pm$0.32 & & 0 & 9.95 \\ 
Lid$\_$635 & 2.01 & 6$\pm$3 & 2$\pm$3 & 7$\pm$4 & 16$\pm$5 & 0.71$\pm$0.6& yes & 0 & 3.43 \\ 
Cid$\_$1512 & 2.02 & 9$\pm$4 & 0$\pm$2 & 11$\pm$4 & 20$\pm$5 & 0.73$\pm$0.56& yes & 56.4 & 56.40 \\ 
Cid$\_$282 & 2.02 & 13$\pm$4 & 11$\pm$4 & 6$\pm$3 & 30$\pm$6 & 0.18$\pm$0.28&  & 0 & 0.79 \\ 
Cid$\_$351 & 2.02 & 54$\pm$8 & 20$\pm$5 & 11$\pm$4 & 86$\pm$10 & -0.11$\pm$0.13 & & 0 & 2.41 \\ 
Cid$\_$424 & 2.02 & 8$\pm$4 & 6$\pm$3 & 7$\pm$3 & 22$\pm$5 & 0.46$\pm$0.41 & yes & 5.65 & 0.79 \\ \hline \hline
\label{table:COSMOS}
\end{tabular}
\end{center}

\flushleft \footnotesize{$^{a}$Identification number of the source, from the COSMOS legacy survey \citep{Marchesi2016}.}\\
\footnotesize{$^{b}$Spectroscopic redshift from \cite{Marchesi2016}.}\\
\footnotesize{$^{c}$Number of counts in the soft energy range $\mathrm{[8-12]/(1+z)}$ keV.}\\
\footnotesize{$^{d}$Number of counts in the mid energy range $\mathrm{[12-16]/(1+z)}$ keV.}\\
\footnotesize{$^{e}$Number of counts in the hard energy range $\mathrm{[16-24]/(1+z)}$ keV.}\\
\footnotesize{$^{f}$Number of counts in the total energy range $\mathrm{[8-24]/(1+z)}$ keV.}\\
\footnotesize{$^{g}$Measured Spectral Curvature values.}\\
\footnotesize{$^{h}$Column density from \cite{Marchesi2016} estimated using a hardness ratio.}\\
\footnotesize{$^{i}$Column density estimated from \cite{Marchesi2016a} spectral fitting.}\\

\label{}
\end{table*}


\section*{Acknowledgements}
M. K. acknowledges support from the SNSF through the Ambizione fellowship grant PZ00P2\textunderscore154799/1.  M.K. and K. S. acknowledge support from  Swiss National Science Foundation Grants PP00P2\_138979 and PP00P2\_166159. The scientific results reported in this article are based on data obtained from the \textit{Chandra} Data Archive.

This research made use of {\tt Astropy}, a community-developed core Python package for Astronomy (Astropy Collaboration, 2013) and the NASA's Astrophysics Data System.



\bibliographystyle{mn2e}
\bibliography{SC_Chandra} 


\bsp	
\label{lastpage}
\end{document}
